\documentclass[apjfonts]{emulateapj}
\usepackage{graphicx}

\DeclareMathAlphabet{\mathbi}{\encodingdefault}{\rmdefault}{\bfdefault}{\itdefault}
\renewcommand\vec[1]{\ensuremath{\mathbi{#1}}}

\begin{document}
\title{Measuring the mass of solar system planets using pulsar timing}

\author{D. J. Champion\altaffilmark{1,2}, G. B. Hobbs\altaffilmark{1},
  R. N. Manchester\altaffilmark{1}, R. T. Edwards\altaffilmark{1,3},
  D. C. Backer\altaffilmark{4}, M. Bailes\altaffilmark{5},
  N. D. R. Bhat\altaffilmark{5}, S. Burke-Spolaor\altaffilmark{5,1},
  W. Coles\altaffilmark{6}, P. B. Demorest\altaffilmark{7},
  R. D. Ferdman\altaffilmark{8}, W. M. Folkner\altaffilmark{9},
  A. W. Hotan\altaffilmark{10}, M. Kramer\altaffilmark{2},
  A. N. Lommen\altaffilmark{11}, D. J. Nice\altaffilmark{12},
  M. B. Purver\altaffilmark{8}, J. M. Sarkissian\altaffilmark{1},
  I. H. Stairs\altaffilmark{13}, W. van Straten\altaffilmark{5},
  J. P. W. Verbiest\altaffilmark{2}, D. R. B. Yardley\altaffilmark{14,1}}

\altaffiltext{1}{CSIRO Astronomy
  and Space Science, Australia Telescope National Facility, P.O. Box 76, Epping, 
NSW 1710, Australia}
\altaffiltext{2}{Max-Planck-Institut f\"{u}r Radioastronomie, Auf dem H\"{u}gel 69, 
53121 Bonn, Germany}
\altaffiltext{3}{The Kilmore International School, 40 White St,
  Kilmore, Victoria 3764, Australia}
\altaffiltext{4}{Department of Astronomy and Radio Astronomy Laboratory, 
University of California, Berkeley, CA 94720, USA}
\altaffiltext{5}{Swinburne University of Technology, P.O. Box 218, Hawthorn, 
Victoria 3122, Australia}
\altaffiltext{6}{Electrical and Computer Engineering, University of California at San Diego,
 La Jolla, CA, USA}
\altaffiltext{7}{National Radio Astronomy Observatory, Charlottesville, VA 22901, USA}
\altaffiltext{8}{Jodrell Bank Centre for Astrophysics, University of Manchester, 
 Manchester, M13 9PL, UK}
\altaffiltext{9}{Jet Propulsion Laboratory, California Institute of Technology, 4800 Oak Grove Dr, Pasadena, CA 91109-8099, USA}
\altaffiltext{10}{Curtin Institute of Radio Astronomy, Curtin
  University, Bentley, WA 6102, Australia}
\altaffiltext{11}{Franklin and Marshall College, 415 Harrisburg Pike, Lancaster, 
PA 17604, USA}
\altaffiltext{12}{Physics Department, Lafayette College, Easton, PA 18042, USA}
\altaffiltext{13}{Department of Physics and Astronomy, University of British Columbia, 
6224 Agricultural Road, Vancouver, BC V6T 1Z1, Canada}
\altaffiltext{14}{Sydney Institute for Astronomy, School of Physics, The University 
 of Sydney, NSW 2006, Australia}
 
\email{champion@pulsarastronomy.net}

\begin{abstract}
  High-precision pulsar timing relies on a solar-system ephemeris in
  order to convert times of arrival (TOAs) of pulses measured at an
  observatory to the solar system barycenter. Any error in the
  conversion to the barycentric TOAs leads to a systematic
  variation in the observed timing residuals; specifically, an
  incorrect planetary mass leads to a predominantly sinusoidal
  variation having a period and phase associated with the planet's
  orbital motion about the Sun. By using an array of pulsars (PSRs
  J0437$-$4715, J1744$-$1134, J1857$+$0943, J1909$-$3744), the masses
  of the planetary systems from Mercury to Saturn have been
  determined. These masses are consistent with the best-known masses
  determined by spacecraft observations, with the mass of the Jovian
  system, 9.547921(2)$\times 10^{-4}$\,M$_\odot$, being significantly
  more accurate than the mass determined from the \emph{Pioneer} and
  \emph{Voyager} spacecraft, and consistent with but less accurate
  than the value from the \emph{Galileo} spacecraft. While spacecraft
  are likely to produce the most accurate measurements for individual
  solar system bodies, the pulsar technique is sensitive to planetary
  system masses and has the potential to provide the most accurate
  values of these masses for some planets.
\end{abstract}


\section{Introduction}
The technique of pulsar timing can provide precise measurements of the
rotational, astrometric, and orbital parameters of a pulsar by modeling
the observed pulse times of arrival (TOAs). The basic timing analysis
provides a fittable parametric model of delays associated with
variations in the Euclidean distance between the pulsar and the Earth
(resulting from Earth's orbital motion, the proper motion of the
pulsar, and its binary motion), dispersive delays in the interstellar
medium, and general relativistic time dilation of clocks in the
observatory and pulsar frames and along the propagation path
\citep[see, e.g.,][]{ehm06}. The largest variable delay term is the
so-called Roemer delay: the modulation caused by the orbital motion of the
Earth relative to the solar system barycenter (SSB). The amplitude of
this delay is up to $\sim$500\,s, while pulse TOAs for many pulsars
are measurable with an uncertainty of much less than 1~$\mu$s. This
delay is compensated using a numerical solar system ephemeris
\citep[e.g.,][]{sta98b}. However, the solar system ephemerides cannot
be perfect and, at some level, will introduce systematic effects into
the timing process. In addition to their use in pulsar timing, these
ephemerides are used to provide guidance information for space
missions (in fact, this was the original motivation for their
development), and hence there is considerable interest in improving
their accuracy.

The measured TOAs, $t_i$, are related to the rotational phase,
$\phi_i$, of the pulsar at the time of emission as follows:
\begin{equation}\label{eq:phase}
  \phi_i = \nu {\textsc t}_i + \frac{\dot\nu {\textsc t}_i^2}{2} + \ldots,
\end{equation}
where
\begin{equation}\label{eq:time}
  {\textsc t}_i = t_i + \frac{(\vec{s_i}+\vec{r_i})\cdot\vec{R}}{c} - \Delta_i. 
\end{equation}
Here, ${\textsc t}_i$ is the time of pulse emission, $\vec{s_i}$ and
$\vec{r_i}$ are, respectively, the vectors from the SSB to the geocenter
and from the geocenter to the observatory at time $t_i$, and $\vec{R}$
is a unit vector from the SSB toward the pulsar. $\Delta_i$ accounts
for numerous other delays not relevant to the present discussion
\citep[see, e.g.,][]{ehm06}. Equation~(\ref{eq:phase}) expresses the
rotational behavior of the pulsar as a Taylor series, which for most
millisecond pulsars receives significant non-stochastic contributions
from only the two terms shown. Equation~(\ref{eq:time}) relates the
times of emission and reception, explicitly including the variations
in light-travel time resulting from the motion of the observatory with
respect to the SSB. If the parameters of the timing model are perfect,
then $\phi_i$ is always an integer. The differences between the
observed phase and that predicted by the timing model are referred to
as the ``timing residuals'', usually expressed in time units through
division by $\nu$. The best-fit timing model is generally that which
minimizes the weighted sum of the squared residuals, where the weights
are the reciprocals of the squared measurement uncertainties in $t_i$.

The most commonly used solar system ephemerides for pulsar timing are
from NASA's Jet Propulsion Laboratory (JPL). They are constructed by
numerical integration of the equations of motion and adjustment of the
model parameters to fit data from optical astrometry, astrolabe
measurements, observations of transits and occultations of the planets
and their rings, radar ranging of the planets, radio astrometry of the
planets using very long baseline interferometry, radio ranging and
Doppler tracking of spacecraft, and laser ranging of the Moon
\citep{sta98b}. These observations constrain the motion of solar system bodies
with respect to the Earth, however they do not
tightly constrain the planetary masses. This is reflected in the fact
that the planetary/solar mass ratios are normally held fixed in the
fit.

If the vector between the observatory and SSB is not correctly
determined, then systematic timing residuals will be induced. For
instance, if the mass of the Jovian system is in error, then
sinusoidal timing residuals with a period equal to Jupiter's orbital
period will be induced. The identification of such residuals therefore
provides a method to limit or detect planetary mass errors in the
solar system ephemeris.

In this letter we use data taken as part of the international effort to
detect gravitational waves \citep{man08,jsk+08,jfl+09,haa+10} using an
array of pulsars to constrain the masses of the solar system planetary
systems. These data sets are described in Section~\ref{ch:DataSets} and
their analysis is discussed in Section~\ref{ch:Analysis}. In
Section~\ref{ch:Results} the results are presented and in Section~\ref{ch:Disc} we
discuss the potential of future observations and the constraints on
unknown solar system bodies.

\section{Data sets}
\label{ch:DataSets}
The pulsars used in this analysis (listed in Table~\ref{tab:data})
were chosen from the sample observed as part of the International
Pulsar Timing Array project \citep{haa+10}. The four pulsars
were selected based upon the precision of their measured TOAs, the
magnitude of timing irregularities and on the length of the data
set. The data sets for PSRs~J0437$-$4715, J1744$-$1134, and
J1909$-$3744 are those published by \citet{vbc+09} except for a
reweighting as described below.

\begin{deluxetable}{llrrc}
\tablecaption{The Data Sets}
\tablecolumns{3}
\tablehead{
\colhead{Name} & \colhead{MJD Range} & \colhead{Years} & \colhead{TOAs} & \colhead{Rms Residual ($\mu$s)}}
\startdata
J0437$-$4715 & 50190 -- 53819 &  9.9 & 2847 & 0.21\\
J1744$-$1134 & 49729 -- 54546 & 13.2 &  342 & 0.64\\
J1857$+$0943 & 46436 -- 54507 & 22.1 &  592 & 1.34\\
J1909$-$3744 & 52618 -- 54528 &  5.2 &  893 & 0.17
\enddata
\label{tab:data} 
\end{deluxetable}

For each observation of a pulsar, typically of 1 hr duration, the
data are folded at the rotation period of the pulsar and summed to
produce a single pulse profile of relatively high signal-to-noise
ratio. The TOA for each profile was obtained by cross-correlating the profile
with a high signal-to-noise ratio template and adjusting the start time
of the observation for the phase offset between the template and
observed profiles. The \textsc{psrchive} \citep{hvm04} and
\textsc{tempo2} \citep{hem06} software packages were used to process
the data and to obtain timing solutions.

The data set for PSR~J1857$+$0943 is a combination of the previously
published TOA data from the Arecibo telescope \citep{ktr94} in
addition to new data from Arecibo, Parkes, and Effelsberg. This
combined data set is over 22 years long. Even though this data set is
nearly 10 years longer than the other data sets in our sample,
accurate TOAs were not obtained for just over 3 years during the
upgrade of the Arecibo telescope. The lack of useful data means that
an arbitrary offset has to be included between the pre- and
post-upgrade data sets. This arbitrary offset absorbs low-frequency
power in the residuals which reduces the sensitivity of the fit to
low-frequency terms.

The uncertainties for the parameters produced by the standard weighted
least-squares fit implemented into \textsc{tempo2} assume that the
reduced $\chi^2$ of the fit is unity. In most pulsar data sets the
reduced $\chi^2$ of the fit is significantly larger than one. There
are a number of possible reasons for this, including: radio frequency
interference causing subtle shape changes in the profile, variations
in the interstellar propagation path, intrinsic variations in the
pulse profile or the pulsar rotation rate, instrumental artifacts,
errors in the clocks used to timestamp the data, or gravitational
waves. Many of these effects have a steep-spectrum or ``red''
character and manifest approximately as low-order polynomials in the
timing residuals. In order to improve the estimate of the TOA
uncertainty (and therefore the uncertainty of the parameters in the
fit), it is common practice to introduce a multiplier that is applied
to the TOA uncertainties at fitting. This is usually determined by
fitting a polynomial to ``whiten'' (i.e., flatten) the residuals and
then calculating the multiplier required to bring the reduced $\chi^2$
to unity. Because of the ad hoc nature of this process and because we
are searching for long-period signals (i.e., signals with periods
similar to the length of our data sets), we use an improved technique
to whiten the data and obtain accurate timing model parameters in the
presence of red noise and with poorly known TOA uncertainties. This
technique is called ``Cholesky whitening'' and is summarized briefly
in Section 3, but will be described more fully in an upcoming paper.

\section{Analysis}
\label{ch:Analysis}
\begin{figure}
\includegraphics [scale=0.35,angle=-90] {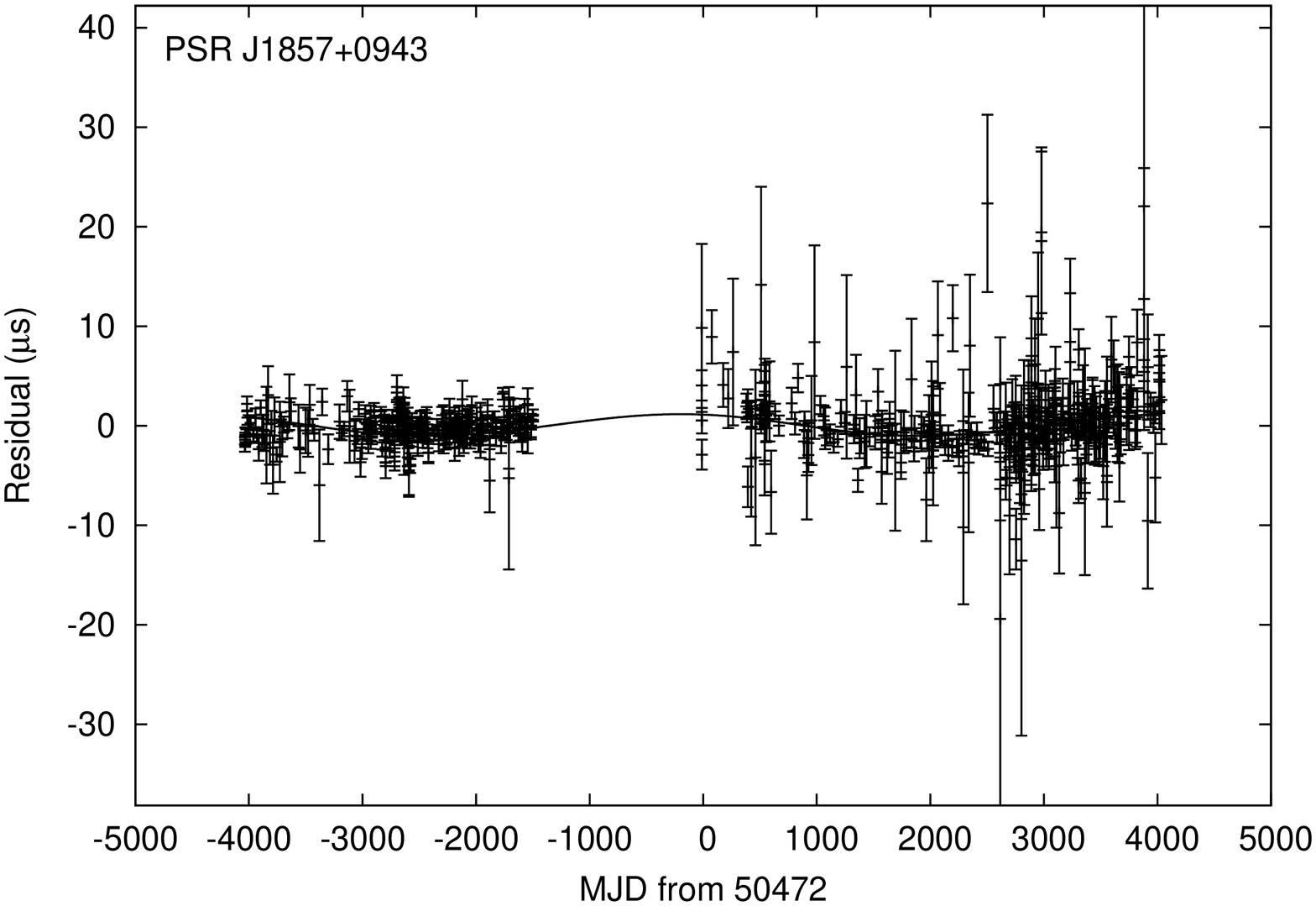}
\includegraphics [scale=0.35,angle=-90] {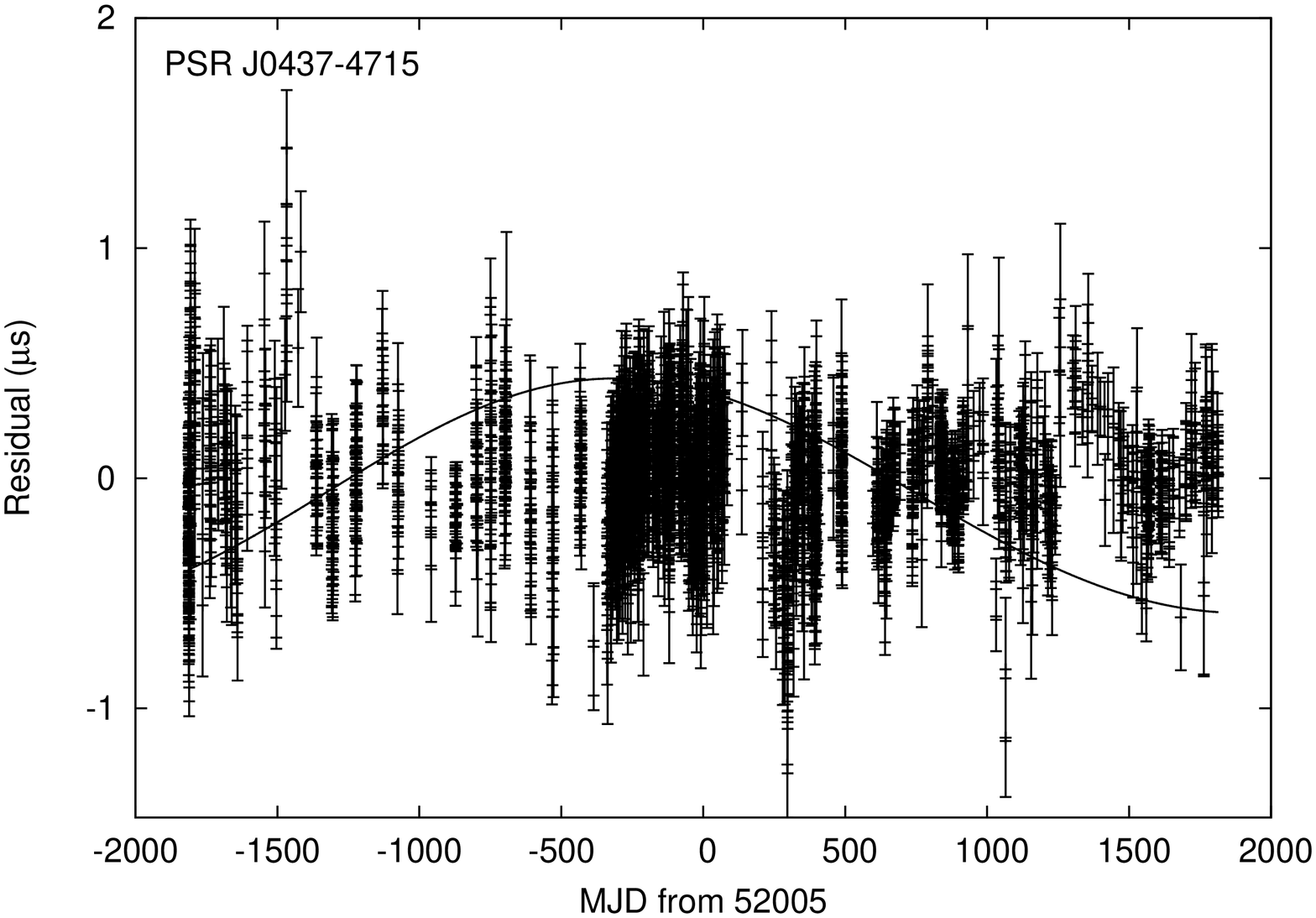}
\caption{Timing residuals for PSRs J1857$+$0943 and J0437$-$4715
  using the DE421 ephemeris plotted with the line indicating the timing
  signature generated by an increase in the mass of Jupiter of
  $5\times10^{-10}$ M$_\odot$.}\label{fig:resids}
\end{figure}

The position of the SSB in a Euclidean frame can be written as a sum
over all solar system bodies (including the Sun), where $M_j$ is the
mass of the body and $\vec{b_j}$ the vector position of the body
(where the $i$ subscripts used in Equations (1) and (2) have been dropped for clarity):
\begin{equation}
\vec{b_B} = \sum_j \vec{b_j}\frac{M_j}{M_T},
\end{equation}
where $M_T = \sum_j M_j$.
An erroneous set of masses $M'_j = M_j-\delta_j$ leads to an erroneous
estimate of the barycenter vector
\begin{equation}
\vec{b'_B} - \vec{b_B} \approx -\sum_j \vec{b_j}\frac{\delta_j}{M_T},
\label{eq:poseffect}
\end{equation}
where it has been assumed that $\delta_j \ll M_j$ and, consequently, that $\sum_j M'_j
\approx \sum_j M_j$.  If we take the origin of the reference frame to
be at the SSB, then $\vec{b_B} = 0$ and
\begin{equation}
\vec{s'} - \vec{s} = -\vec{b'_B} \approx
\sum_j \vec{b_j}\frac{\delta_j}{M_T}. 
\end{equation}
The error in the model time of emission is then
\begin{eqnarray}
{\textsc t}' - {\textsc t} &=& \frac{\left(\vec{s}' - \vec{s}\right) 
\cdot \vec{R}}{c} \\
 &\approx& \frac{1}{cM_\odot} \sum_j \delta_j (\vec{b_j} \cdot \vec{R}),
\label{eq:timingeffect}
\end{eqnarray}
that is, we approximate the effect of a change in the mass of a planet
as a relocation of the SSB along the vector from the original SSB to
that planet. The \textsc{tempo2} software
package has been modified to include the right side of
Equation~(\ref{eq:timingeffect}) as additional terms in the model. The
modified timing model obtains $\vec{b_j}$ from a specified version of
the JPL series of solar system ephemerides \citep[in this work,
DE421,][]{fwb09}. The model parameters $\delta_j$ measure the
difference between the best-fit masses and the values assumed by the
chosen solar system ephemeris. Indices $j$ of 1 -- 9 refer to the
planets (and Pluto) in ascending order of mean distance from the Sun
(note that $\vec{b_3} \equiv \vec{s}$). Examples of the induced timing
residuals resulting from an increase in the Jovian system mass of
$5\times10^{-10}$ M$_\odot$ are given in Figure~\ref{fig:resids} for
PSRs J0437$-$4715 and J1857$+$0943 (ignoring any effects caused by the
fitting procedures). 

In order to deal with all our data sets, we adapted \textsc{tempo2} to
fit multiple pulsars simultaneously. Fitting for the pulsar specific
parameters is based solely on the TOAs of that pulsar, whereas
pulsar-independent parameters, such as a planetary mass, are fitted
globally over all data sets. This procedure reduces the impact of
timing noise in individual pulsars on the derived values for the
global parameters.

For data sets whose post-fit residuals have a reduced $\chi^2$
value close to 1.0, it is possible simply to fit for the planetary system mass. 
To determine realistic uncertainties for the TOAs, we selected
short sections of data ($\sim 30$ days long depending upon the sampling)
for each pulsar, observatory, and back-end instrument used. Weighted
fits to each of these short sections of data gave independent
estimates of the correction factors which were subsequently averaged
and applied to the data set for that combination. This procedure
avoided contamination of the correction factors by non-white noise in
the data sets. In the presence of non-white noise, standard fitting
procedures lead to biased parameter estimates and underestimated
uncertainties \citep[see, e.g.,][]{vbv+08}. 

The red noise in each data set was modeled following the method
outlined by \citet{vbv+08}. For PSR~J0437$-$4715 the model developed
by \citet{vbv+08} was used, while for PSRs J1744$-$1134, J1857$+$0943,
and J1909$-$3744, the red noise was fitted by a power-law function, $P
\propto f^{-\alpha}$, with exponents of $\alpha = 0.9$, $0.7$, and
$0.55$, respectively. These noise models provide two methods to
determine the parameter uncertainties. First, conservative estimates
of the parameter uncertainties were obtained using a Monte-Carlo
simulation as described by \citet{vbv+08}. Second, we implemented a
new technique that both whitens the residuals and modifies the function
being fitted before obtaining the parameter values and uncertainties
using a Cholesky factorization of the data covariance matrix. This
procedure, known as ``Cholesky whitening'', will be fully described in
a forthcoming paper.

To test our analysis technique, a new ephemeris was created that had
identical parameters to the DE421 ephemeris, except for a small
decrease in the mass of Jupiter by $7\times10^{-11}$ M$_{\odot}$. The
effect of this change was investigated by simulating TOAs that are
predicted exactly by a given timing model and the DE421
ephemeris. These simulated data were then analyzed using {\sc tempo2}
with the modified ephemeris. The resulting pre-fit residuals show the
expected sinusoid at Jupiter's orbital period together with an annual
term of about half the amplitude of the Jupiter term. Changing the
mass of Jupiter has many secondary effects in the modified
ephemeris. These include a slight variation in the Astronomical Unit
which leads to the annual sinusoid. This small effect is likely to be
undetectable in real data and in any case would be absorbed as an
offset in the position of the pulsar by $\sim 0.1$
mas. The {\sc tempo2} fitting correctly recovered the
simulated offset in Jupiter's mass.

\section{Results}
\label{ch:Results}

Using the DE421 ephemeris, we have obtained timing residuals for the
four pulsars listed in Table~\ref{tab:data} and fitted for a mass
difference for each of the planetary systems from Mercury to
Saturn. The resulting mass measurements are listed in
Table~\ref{tab:results}, where the 1-$\sigma$ uncertainties given in
parentheses are in the last quoted digit. All results from this work
are consistent with the best current measurements; the number of
standard deviations between the masses derived in this work and the
best-known masses are given in last column.

\begin{deluxetable}{llcll}
\tablecaption{Planetary system masses}
\tablecolumns{3}
\tablehead{
\colhead{System} & \colhead{Best-Known Mass (M$_\odot$)}& \colhead{Ref.} & \colhead{This Work (M$_\odot$)} & \colhead{$\delta_j/\sigma_j$} }
\startdata
Mercury & 1.66013(7)$\times 10^{-7}$     & 1 & 1.6584(17)$\times 10^{-7}$  & 1.02\\
Venus   & 2.44783824(4)$\times 10^{-6}$  & 2 & 2.44783(17)$\times 10^{-6}$ & 0.05\\
Mars    & 3.2271560(2)$\times 10^{-7}$   & 3 & 3.226(2)$\times 10^{-7}$    & 0.58\\
Jupiter & 9.54791898(16)$\times 10^{-4}$ & 4 & 9.547921(2)$\times 10^{-4}$ & 1.01\\
Saturn  & 2.85885670(8)$\times 10^{-4}$  & 5 & 2.858872(8)$\times 10^{-4}$ & 1.91
\enddata
\tablerefs{(1) \citet{ace+87}; (2) \citet{kbs99a}; (3) \citet{kys+06}; (4) \citet{jhm+00}; (5) \citet{jab+06}.}
\label{tab:results} 
\end{deluxetable}

The mass measurement for Mars was determined without the use of the
PSR~J0437$-$4715 data. A spectral analysis of the data set shows
significant power in a broad feature around the period of the Martian
orbit which could contaminate a fit for the narrow feature that would
indicate an error in the mass of Mars. The simple red-noise model used
to calculate the correct uncertainties is not detailed enough to
account for this feature and so this data set was not used.

\begin{figure}
\includegraphics [scale=0.35,angle=-90] {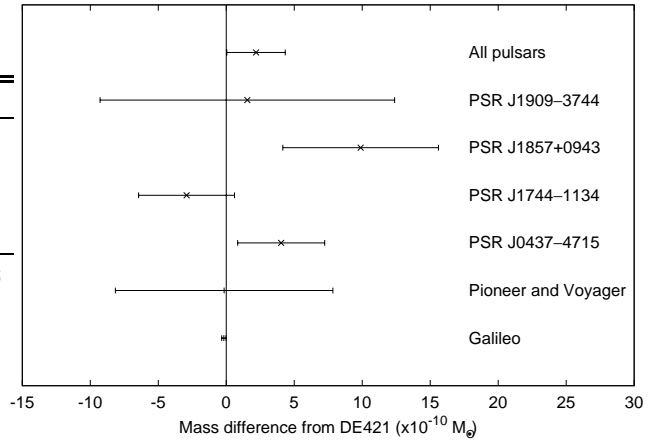}
\caption{Values and uncertainties for the mass of the Jovian system
  from the \emph{Pioneer} and \emph{Voyager} \citep{cs85}, and the
  \emph{Galileo} \citep{jhm+00} spacecraft , the pulsars individually,
  and the array of pulsars.}
\label{fig:jupmass}
\end{figure}

Our current data sets are sensitive to mass differences of
approximately $10^{-10}$\,M$_\odot$, independent of the
planet. Consequently, our most precise fractional mass determination
is for the Jovian system. We therefore check our result by comparing
the Jovian system mass obtained using different subsets of our
data. In Figure~\ref{fig:jupmass}, we show the fitted mass difference
compared with the value used for the DE421 ephemeris,
$9.5479191563\times 10^{-4}$~M$_\odot$, for each pulsar separately and
the weighted mean.  For comparison, we also show the best Jovian system
mass from the \emph{Pioneer} and \emph{Voyager} \citep{cs85} and the
\emph{Galileo} \citep{jhm+00} spacecraft. The results obtained by
fitting to individual pulsar data sets show a small scatter around
the DE421 mass value with no pulsar showing more than a 2-$\sigma$
deviation. The weighted mean deviates from the best-known measurement
by only 1.1\,$\sigma$ and has considerably smaller uncertainties than
the mass determination derived from \emph{Pioneer} and
\emph{Voyager} \citep{cs85}.

\section{Discussion}
\label{ch:Disc}

\begin{figure}
\includegraphics [scale=0.35,angle=-90] {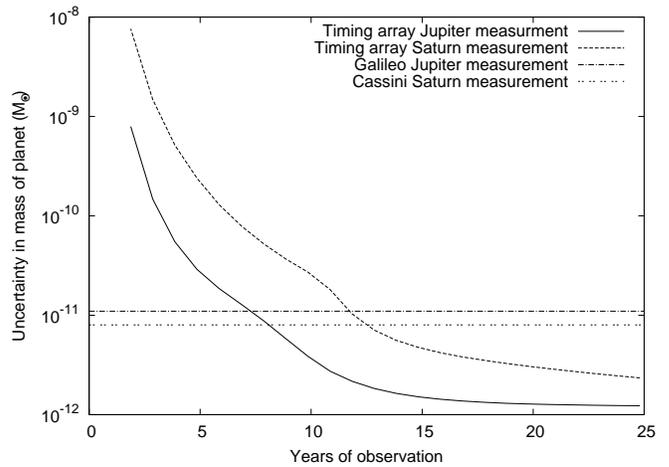}
  \caption{Mass uncertainties for the Jovian and Saturnian system
    using simulated data from an array of 20 pulsars sampled
    every 14 days, timed with an rms timing residual of 100\,ns for
    different data spans. Also plotted are the current best mass
    measurements for Jupiter \citep{jhm+00} and Saturn \citep{jab+06}.}
\label{fig:array}
\end{figure}

While the result presented here for the Jovian system is more precise
than the best measurement derived from the \emph{Pioneer} and
\emph{Voyager} spacecraft by a factor of $\sim$4, the result from the
\emph{Galileo} spacecraft is still better by a factor of $\sim
20$. For a pulsar timing array of 20 pulsars, regularly sampled every
two weeks, with white data sets and an rms timing residual of 100\,ns,
the uncertainty of the mass estimate decreases with increasing data
span such that the mass uncertainty of the \emph{Galileo} measurement
for Jupiter would be reached in $\sim 7$ years of observations; see
the solid line in Figure~\ref{fig:array}. Note that this curve does
not follow a simple power-law function because of the fitting
procedures that are undertaken when dealing with pulsar data
sets. Figure~\ref{fig:array} also shows that, with $\sim$13\,years of
data, the uncertainty of the current \emph{Cassini} measurement for
Saturn is reached.

These predictions rely on pulsar data sets remaining ``white'' over
timescales of a decade or more at very high levels of timing
precision. While this has yet to be demonstrated, the indications from
recent decade-long data sets \citep{vbc+09} are encouraging.

Analysis of data from current and future spacecraft will produce
improved measurements of planetary masses. For example, NASA's New
Frontiers Mission to Jupiter, \emph{Juno}, is expected to reach
Jupiter in 2016 and orbit it for more than a year. A major scientific
objective of this mission is to probe Jupiter's gravitational field in
detail, thereby providing a very precise mass for Jupiter.

While spacecraft measurements are likely to continue to provide the
most precise mass measurements for most of the planets, at least for
the next decade, it should be noted that the pulsar measurements are
independent with different assumptions and sources of uncertainty.
Independent methods are particularly important for high-precision
measurements where sources of systematic error may not be well
understood.  Furthermore, while spacecraft such as \emph{Juno} are very
sensitive to the mass of the individual body being orbited, they tell
us very little about the satellite masses of that body. Only five of
the Jovian and nine of the Saturnian satellites are included in the
system mass assumed for DE421 (R. A. Jacobson, private
communication). Since the pulsar technique is sensitive to the mass of
the entire planetary system, it can provide a measure of the mass
undetermined by spacecraft observations.

By combining the pulsar and satellite measurements, it will be possible
to test the inverse-square relation of gravity and distance for
Jupiter masses and distances between 0.1 and 5~AU. Although no
deviations apart from known general relativistic effects are expected,
it is important to place limits on such effects where possible.

The pulsar timing technique is also sensitive to other solar system
objects such as asteroids and currently unknown bodies, e.g.,
trans-Neptunian objects (TNOs). Measurements of anomalous period
derivatives and binary period derivatives for a number of millisecond
pulsars have already been used to place limits on the acceleration of
the Solar System toward nearby stars or undetected massive planets
\citep[e.g.,][]{zt05,vbv+08}. Pulsar timing array experiments with a
wide distribution of pulsars on the sky will be sensitive to the
dipolar spatial dependence resulting from any error in the
solar system ephemeris, including currently unknown TNOs. Any
ephemeris error will be distinguishable from the effects of
gravitational waves passing over the Earth as the latter have a quadrupolar
spatial signature. Limits for unknown masses have also been placed by
spacecraft using deviations from their predicted trajectories. Doppler
tracking data from the two \emph{Pioneer} spacecraft were searched for
accelerations due to an unknown planet. The anomalous acceleration
detected in these data, $a_P = (8.7\pm 1.3) \times 10^{-10}$
m~s$^{-1}$ is attributed to non-gravitational sources \citep{all+02}
and is not detected in planetary measurements \citep{fol10}.

An exciting possibility for the future is the creation of a
solar system ephemeris that includes pulsar timing data in the overall
fit. Such a fit would be able to determine the masses of the planetary
systems while simultaneously fitting for orbital parameters.

\section{Conclusions}
We have used the four longest and most precise data sets taken for
pulsar timing array projects to constrain the masses of the
solar system planetary systems from Mercury to Saturn. In all cases,
these measurements are consistent with the best-known
measurements. For the Jovian system, our measurement improves on the
\emph{Pioneer} and \emph{Voyager} spacecraft measurement and is
consistent with the mass derived from observations of the
\emph{Galileo} spacecraft as it orbited the planet between 1995 and
2003. Pulsar timing has the potential to make the most accurate
measurements of planetary system masses and to detect currently
unknown solar system objects such as TNOs. In the future, pulsar
timing data can be included in the global solutions used to derive
solar system ephemerides, thereby improving their precision.

\section*{Acknowledgments}
We thank R. A. Jacobson for useful information about the planetary
mass determinations, and F. A. Jenet, S. Oslowski, E. Splaver,
K. Xilouris and X. P You for assistance with the observations and
analysis. We also thank the referee for constructive comments. This work is
undertaken as part of the Parkes Pulsar Timing Array project which was
supported by R.N.M.'s Australian Research Council Federation Fellowship
(FF0348478). G.H. is the recipient of an Australian Research Council
QEII Fellowship (DP0878388), A.L. is the recipient of a National
Science Foundation CAREER award (AST-0748580) and D.N. the recipient of an NSF grant (AST-0647820). Pulsar research at UBC
is funded by an NSERC Discovery Grant. The Parkes radio telescope is
part of the Australia Telescope, which is funded by the Commonwealth
of Australia for operation as a National Facility managed by the
Commonwealth Scientific and Industrial Research Organisation. Part of the research described in this paper was carried out at the Jet Propulsion
 Laboratory, California Institute of Technology, under contract with the
 National Aeronautics and Space Administration.

\end{document}